\documentclass[letterpaper,10pt]{article} 

\usepackage{opticameet3} 
\usepackage{tikz}
\usepackage{pgfplots}
\pgfplotsset{scaled y ticks=false}

\newcommand\authormark[1]{\textsuperscript{#1}}

\usepackage{amsmath,amssymb}
\usepackage[colorlinks=true,bookmarks=false,citecolor=blue,urlcolor=blue]{hyperref} 

\begin{document}

\title{Optimization of Pilot-Aided Joint Phase Recovery for Frequency Comb-Based Wideband Transmission} 

\author{Gabriele Di Rosa\authormark{1,*}, Ognjen Jovanovic\authormark{1}, M. Ahmed Leghari\authormark{1}, Jasper M\"uller\authormark{1}, Benjamin Wohlfeil\authormark{2}, and J\"org-Peter Elbers\authormark{1}}

\address{\authormark{1} Adtran Networks SE, Fraunhoferstr. 9a, 82152 Martinsried/Munich, Germany\\
\authormark{2} Adtran Networks SE, Justus-von-Liebig Str. 7, 12489 Berlin, Germany}

\email{\authormark{*}gabriele.dirosa@adtran.com} 

\begin{abstract}
We numerically investigate joint pilot-aided phase recovery for frequency comb-based long-haul wideband transmission. We report net information rate gains by optimizing the pilot overhead and phase estimation algorithm, outperforming per-channel processing at lower complexity.
\vspace{-1\baselineskip}
\end{abstract}

\section{Introduction}
\vspace{-0.5\baselineskip}
In the last decade, advancements in CMOS processes and high-bandwidth optical components \cite{thin_film_mod} have enabled conventional transmission with a single optical carrier to drive the reduction of the cost-per-bit. Nevertheless, it is unclear how long this approach will be sustainable. Optical frequency combs (OFCs) are a promising solution for ultra-high capacity communication systems exploiting parallel signal generation and detection \cite{superchannel2015}.

Multi-wavelength concepts are increasingly considered an option for developing scalable integrated transceivers \cite{ScalingIntegrationChallenges} and the potential of OFCs as multi-wavelength sources for coherent transmission has been demonstrated \cite{schroder2019laser, marin2020performance, jorgensen2022petabit,jasperJOCN}. Features of OFCs, such as fixed spacing and coherence among the optical carriers, enable joint carrier recovery (CR) \cite{alfredsson2019performance,lundberg2020phase,lundbergMDPIreview}. This aspect is advantageous for reducing the complexity of digital signal processing (DSP) algorithms, especially for shorter transmission reach and higher order modulation, where CR has a higher relative weight in the overall complexity \cite{ChalmersPowerConsumption}. Simple approaches, such as assuming perfect coherence between the different carriers and estimating the phase noise (PN) on one signal only, become less effective with large accumulated chromatic dispersion (CD) \cite{lundbergMDPIreview}. This is a crucial limitation for next-generation systems based on ultra-high baud rate ($>$130 GBaud) signals. Consequently, more advanced phase estimation schemes have to be considered \cite{alfredsson2019performance,neves2021enhanced}.

Pilot-based CR is simple and robust even for higher-order quadrature amplitude modulation (QAM), making it a practical option for current DSP implementations. Joint processing enables additional optimization of the distribution of the pilot overhead (POH) among different carriers \cite{alfredsson2019performance}. As a result, it is not trivial to find a trade-off between better-performing joint algorithms with higher POH and reduced room for payload allocation.

In this contribution, we numerically evaluate the performance of low-complexity joint CR for an OFC-based system with four 135 GBd signals over approximately 600 GHz bandwidth. We account for realistic transmission impairments and show that optimized joint CR enables net information rate gains for up to 560 km transmission for 64-QAM and 2160 km for 16-QAM over per-channel processing through optimization of the POH.

\vspace{-0.5\baselineskip}
\section{System model}
\begin{figure*}[t]
\centering
\includegraphics[width=0.92\textwidth]{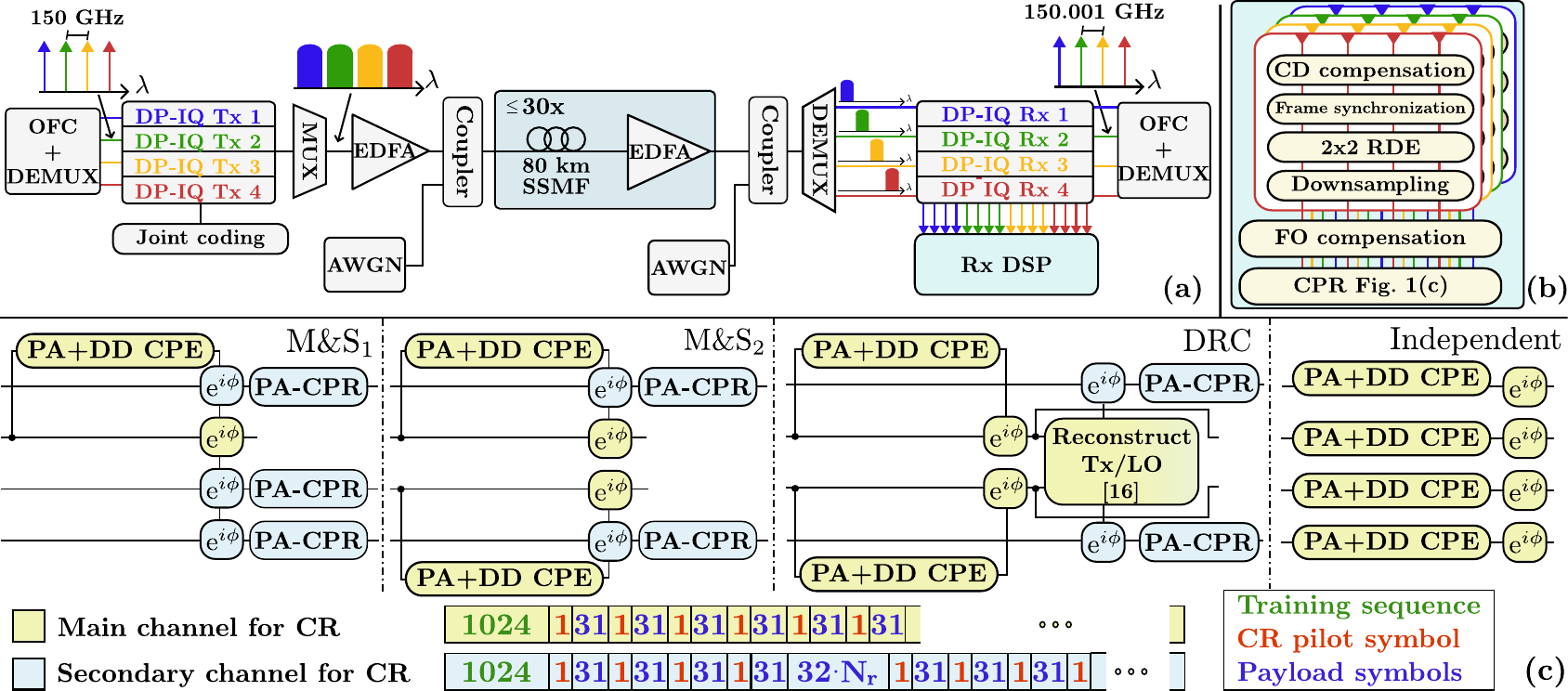}
\caption{(a) Simulation setup (\emph{modeled in VPIphotonics Design Suite 11.2}); (b) DSP chain (\emph{implemented in Python}); (c) considered CR algorithms and pilot symbols distribution schemes.} 
\label{Fig:Setup}
\vspace{-1.5\baselineskip}
\end{figure*}

\vspace{-0.5\baselineskip}
The simulation setup is shown in Fig. \ref{Fig:Setup}(a). OFCs are considered for the transmitter (Tx) lasers and for the local oscillators (LOs) for a four-channel system with 135 Gbaud signals and 150 GHz spacing. The PN of each of the carriers is modeled as the superposition of a common Wiener process described by a $200 \text{ kHz}$ linewidth (LW) and an additional line-dependent PN. The latter is modeled by drawing a single independent Wiener process with $\text{LW} =1 \text{ kHz}$ scaled with the factors [-2,-1,1,2] starting from the leftmost frequency channel. The scaling is used to emulate complete anti-correlation between lines with opposite spacing from the OFC central frequency. This model accurately describes line-dependent phase decorrelation in electro-optic frequency combs \cite{lundberg2020phase}. A frequency offset (FO) of $\approx 200 \text{ MHz}$ between the Tx and LO OFCs is considered, as well as $ \approx 1 \text{ MHz}$ free spectral range (FSR) deviation of the LO from the ideal $150 \text{ GHz}$. This last imperfection is important to validate the robustness of joint CR schemes in the presence of a small time-varying uncalibrated FSR deviation.

Considered modulation formats are 16-QAM and 64-QAM. A header of $1024$ 4-QAM pilot symbols with the same energy as the payload data is inserted at the start of a $2^{17}$ symbols frame. Two successive frames are generated, the signals are pulse shaped with a root-raised-cosine filter with 0.1 roll-off factor and passed to four independent dual-polarization in-phase and quadrature (DP-IQ) modulators. Additive white Gaussian noise (AWGN) after the Tx and before the receiver (Rx) is introduced in equal parts to emulate a back-to-back (B2B) signal-to-noise ratio (SNR) $\text{SNR}_{\text{B2B}} = 22 \text{ dB}$. The signals have a launch power of $3 \text{ dBm}$ per carrier, which is near-optimal for a wide range of propagation distances in the considered scenario. Transmission over up to 30 spans of $80 \text{ km}$ of standard single mode fiber (SSMF) with attenuation, dispersion, and nonlinearity coefficients $\alpha = 0.2 \text{ dB/km}$, $D = 20 \text{ ps/nm/km}$, $\gamma = 1.3 \text{ 1/m/W}$ are simulated. After each span an Erbium doped fiber amplifier (EDFA) with $16 \text{ dB}$ gain and $5.5 \text{ dB}$ noise figure restores the launch power. After fiber propagation the signals are isolated by four optical bandpass filters with a Gaussian profile and a $3 \text{ dB}$ bandwidth of $150 \text{ GHz}$ and received by four coherent receivers with $70 \text{ GHz}$ bandwidth modeled as a $3^{rd}$ order Bessel filter. After analog to digital conversion at 2 samples per symbol, the four channels are processed in the DSP as detailed in Fig. \ref{Fig:Setup}(b): After CD compensation, the 1024 symbols training sequence is exploited for frame synchronization \cite{Park_sync} and pre-convergence of the $2 \times 2$ adaptive equalizer with $31$ fractionally-spaced taps. The equalizer is then blindly adapted according to the radius directed equalization (RDE) scheme. Finally, the signal is downsampled to 1 sample per symbol before CR.

\vspace{-0.5\baselineskip}
\section{Phase recovery and pilot symbols distribution}
\vspace{-0.5\baselineskip}
We consider the four carrier phase recovery (CPR) implementations detailed in Fig. \ref{Fig:Setup}(c). The schemes exploit a variable number of main and secondary channels. For main channels, a PA digital phase locked loop is used to estimate and correct the FO. Carrier phase estimation (CPE) is implemented in two stages. First, PN estimates are averaged over four consecutive pilot symbols to reduce the impact of AWGN. Linear interpolation is performed between consecutive averaged estimates. In the second stage, decision directed (DD) maximum-likelihood phase estimation is performed over an optimized window length to minimize the correlation between the I and Q components \cite{di2020statistical}. We refer to this two-stage scheme as PA+DD CPE. $\text{POH}_{\text{CR}} = 1/31$ is used for the CR pilot symbols, which are distributed as payload data but are known at the Rx. The baseline implementation is represented by independent CR, which has four main channels.

Conventional main and secondary implementations\cite{lundbergMDPIreview} M\&S$_1$ and M\&S$_2$ are based on one and two main channels, respectively. For secondary channels, the FO estimate and CPE from the closest main channel is used. Additionally, an ultra-light pilot-aided carrier phase recovery (PA-CPR) is implemented by estimating and averaging the phase over four consecutive pilot symbols. Unlike for PA+DD CPE, no interpolation is performed but the phase estimated is simply held until the next one to consider a low-latency implementation. 
It is crucial to notice that secondary channels have a reduced $\text{POH}_{\text{CR}} = 1/(31+8\cdot N_r)$, where $N_r$ is the number of consecutive 32-symbol blocks that do not contain pilots. Finally, we consider a dual reference carrier (DRC) transmission-aware method originally devised for digital subcarrier multiplexing. The algorithm exploits the CD estimation and CPE on two carriers to reconstruct the PN components of the Tx laser and LO\cite{neves2021enhanced}. The considered OFC-based system presents additional OFC-induced phase decorrelation and per-channel CPR performance is more robust to laser LW due to a much higher symbol rate per carrier. Therefore, the algorithm is slightly modified such that only the secondary channels phase estimates are derived from the phase reconstruction algorithm. This algorithm has marginally higher complexity than the M\&S$_2$ scheme \cite{neves2021enhanced}.

\vspace{-0.5\baselineskip}
\section{Results}
\begin{figure}[tbp]
    \centering
    \scalebox{0.95}{
%
%
\definecolor{mycolor1}{rgb}{0.00000,0.00000,0.78431}%
\definecolor{mycolor2}{rgb}{0.86275,0.07843,0.23529}%
\definecolor{mycolor3}{rgb}{0.13333,0.54510,0.13333}%
\definecolor{mycolor4}{rgb}{0.9,0.4,0.1}%

\begin{tikzpicture}

\begin{axis}[%
set layers,
width=0.4\linewidth,
height=.3\linewidth,
at={(0.3\linewidth,0\linewidth)},
xmin=80,
xmax=1200,
xlabel style={font=\color{white!15!black}},
xlabel={Transmission distance [km] },
ymin=-0.022,
ymax=0.01,
ylabel style={font=\color{white!15!black}},
ylabel near ticks,
y label style={at={(-0.16,0.5)}},
y tick label style={/pgf/number format/fixed},
title style={},
xmajorgrids,
ymajorgrids,
xtick={80,250,400,600,800,1000,1200}, 
xticklabels = {80,250,400,600,800,1000,1200},
ytick={-0.02,-0.01,0,0.01}, 
yticklabels={-0.02,-0.01,0,0.01}, 
legend columns=1,
legend style={at={(1.6,1)}, anchor=north, legend cell align=left, align=left, draw=white!15!black, font=\footnotesize}
]
\addplot [color=mycolor1,line width = 1pt, dashed, mark=diamond, mark options={solid, mycolor1}]
  table[row sep=crcr]{%
80	-0.008466361\\
160	-0.010583913\\
240	-0.011677992\\
320	-0.012377429\\
400	-0.013165849\\
480	-0.013412165\\
560	-0.013658482\\
640	-0.01442534\\
720	-0.013860663\\
800	-0.013756145\\
880	-0.013252042\\
960	-0.013624632\\
1040	-0.01383336\\
1120	-0.01405548\\
1200	-0.01466295\\
};
\addlegendentry{M\&S$_1$, $N_r = 0$}

\addplot [color=mycolor2,line width = 1pt, dashed, mark=diamond, mark options={solid, mycolor2}]
  table[row sep=crcr]{%
80		-0.0055323\\
160		-0.0068841\\
240		-0.0078560\\
320		-0.0082531\\
400		-0.0088642\\
480		-0.0089302\\
560		-0.0089962\\
640	    -0.0098685\\
720	    -0.0093720\\
800	    -0.0092112\\
880	    -0.0087148\\
960	    -0.0090955\\
1040    -0.0093209\\
1120    -0.0094586\\
1200    -0.0097969\\
};
\addlegendentry{M\&S$_2$, $N_r = 0$}

\addplot [color=mycolor3,line width = 1pt, dashed, mark=diamond, mark options={solid, mycolor3}]
  table[row sep=crcr]{%
80	 	-0.004851917\\
160	    -0.006340124\\
240	    -0.007251236\\
320	    -0.007678033\\
400	    -0.008205107\\
480	    -0.008279881\\
560	    -0.008354656\\
640	    -0.009194057\\
720	    -0.008794578\\
800	    -0.008757338\\
880	    -0.008114917\\
960	    -0.008383477\\
1040    -0.008779807\\
1120    -0.009027818\\
1200    -0.009260294\\
};
\addlegendentry{DRC, $N_r = 0$}

\addplot [color=mycolor1,line width = 1pt, mark=o, mark options={solid, mycolor1}]
  table[row sep=crcr]{%
80	0.002014413\\
160	0.00125848\\
240	-0.000159294\\
320	-0.000555367\\
400	-0.001468833\\
480	-0.002148276\\
560	-0.002827719\\
640	-0.004029636\\
720	-0.004057447\\
800	-0.004328008\\
880	-0.005093296\\
960	-0.004824984\\
1040	-0.006532273\\
1120	-0.006384488\\
1200	-0.006884978\\
};
\addlegendentry{M\&S$_1$, $N_r = 64$}

\addplot [color=mycolor2,line width = 1pt, mark=o, mark options={solid, mycolor2}]
  table[row sep=crcr]{%
80	0.001548991\\
160	0.001468699\\
240	-3.28118E-05\\
320	-0.000331881\\
400	-0.001304274\\
480	-0.001445994\\
560	-0.001587713\\
640	-0.002743079\\
720	-0.002835925\\
800	-0.003264485\\
880	-0.003241945\\
960	-0.003664801\\
1040	-0.004602836\\
1120	-0.004710878\\
1200	-0.004970971\\
};
\addlegendentry{M\&S$_2$, $N_r = 64$}

\addplot [color=mycolor3,line width = 1pt, mark=o, mark options={solid, mycolor3}]
  table[row sep=crcr]{%
80	0.006425507\\
160	0.004528714\\
240	0.002799336\\
320	0.002069782\\
400	0.000768558\\
480	0.000600855\\
560	0.000433153\\
640	-0.00110119\\
720	-0.00121901\\
800	-0.00132382\\
880	-0.00134553\\
960	-0.00193182\\
1040	-0.002361145\\
1120	-0.002430548\\
1200	-0.002997599\\
};
\addlegendentry{DRC, $N_r = 64$}
\addplot [color=black,line width = 1.2pt, dotted]
  table[row sep=crcr]{%
0   0\\
1200   0\\
};
\end{axis}

\begin{axis}[
set layers,
width=0.4\linewidth,
height=.3\linewidth,
at={(0.3\linewidth,0\linewidth)},
xmin=80,
xmax=1200,
xlabel style={font=\color{white!15!black}, at},
ymin=0.74,
ymax=1,
ylabel={\textcolor{mycolor4}{NGMI for independent CPR}},
ylabel near ticks,
y label style={at={(1.2,0.5)}},
axis y line*=right,
axis x line=none,
y tick label style={mycolor4},
y tick style={mycolor4},
title style={at={(0.5,1)}},
title={64-QAM},
ytick={0.75,0.8,0.85,0.9,0.95,1}, 
yticklabels = {0.75,0.8,0.85,0.9,0.95,1}, 
legend style={font=\scriptsize, legend cell align=left, align=left, draw=white!15!black}
]
\addplot [color=mycolor4,line width = 1pt, mark=*, mark size = 1pt]
  table[row sep=crcr]{%
80   0.976062105\\
160	0.960913941\\
240	0.941964324\\
320	0.92245153\\
400	0.902934137\\
480	0.886569983\\
560	0.870205829\\
640	0.850739826\\
720	0.836118353\\
800	0.822519855\\
880	0.806025813\\
960	0.791251621\\
1040 0.776573\\
1120 0.764817\\
1200 0.761974\\
};
\node at (axis cs: 1100,0.97) [anchor=center] {\large (b)};

\end{axis}

\begin{axis}[%
set layers,
width=.4\linewidth,
height=.3\linewidth,
at={(-0.1\linewidth,0\linewidth)},
xmin=400,
xmax=2320,
xlabel style={font=\color{white!15!black}},
xlabel={Transmission distance [km] },
ymin=-0.022,
ymax=0.02,
ylabel style={font=\color{white!15!black}},
ylabel near ticks,
ylabel={$R_{\text{c,net}}$ gain },
y label style={at={(-0.16,0.5)}},
y tick label style={/pgf/number format/fixed},
title style={font=\bfseries},
xmajorgrids,
ymajorgrids,
xtick = {400,800,1200,1600,2000,2320}, 
xticklabels = {400,800,1200,1600,2000,2320},
ytick={-0.02,-0.01,0,0.01,0.02}, 
yticklabels={-0.02,-0.01,0,0.01,0.02},
legend columns=3,
legend style={font=\scriptsize,at={(-0.5,2.57)}, anchor=north, legend cell align=left, align=left, draw=white!15!black}
]
\addplot [color=mycolor1,line width = 1pt, dashed, mark=diamond, mark options={solid, mycolor1}]
  table[row sep=crcr]{%
400	-0.000933099\\
560	-0.001870146\\
720	-0.003472595\\
880	-0.004851244\\
1040	-0.00672575\\
1200	-0.00802853\\
1360	-0.01008047\\
1520	-0.01116070\\
1680	-0.01301840\\
1840	-0.01410085\\
2000	-0.01511253\\
2160	-0.01563389\\
2320	-0.01643524\\
};
\addplot [color=mycolor2,line width = 1pt, dashed, mark=diamond, mark options={solid, mycolor2}]
  table[row sep=crcr]{%
400	-0.000612502\\
560	-0.001208017\\
720	-0.002300494\\
880	-0.003273168\\
1040	-0.00447346\\
1200	-0.00526045\\
1360	-0.00665447\\
1520	-0.00744860\\
1680	-0.00876370\\
1840	-0.00951341\\
2000	-0.00993927\\
2160	-0.01055967\\
2320	-0.01108264\\
};

\addplot [color=mycolor3,line width = 1pt, dashed, mark=diamond, mark options={solid, mycolor3}]
  table[row sep=crcr]{%
400	-0.000576398\\
560	-0.00115608\\
720	-0.002187195\\
880	-0.003117509\\
1040	-0.00422217\\
1200	-0.00506158\\
1360	-0.00638166\\
1520	-0.00710081\\
1680	-0.00837133\\
1840	-0.00908348\\
2000	-0.00949708\\
2160	-0.01015302\\
2320	-0.01062054\\
};

\addplot [color=mycolor1,line width = 1pt, mark=o, mark options={solid, mycolor1}]
  table[row sep=crcr]{%
400	0.01860468\\
560	0.01736001\\
720	0.01539660\\
880	0.01358548\\
1040	0.01108614\\
1200	0.00984972\\
1360	0.00724340\\
1520	0.00575861\\
1680	0.00280928\\
1840	0.00113517\\
2000	-0.0006087\\
2160	-0.0012833\\
2320	-0.0029848\\
};

\addplot [color=mycolor2,line width = 1pt, mark=o, mark options={solid, mycolor2}]
  table[row sep=crcr]{%
400	0.012335378\\
560	0.011494145\\
720	0.010181812\\
880	0.008979939\\
1040	0.007420429\\
1200	0.006409009\\
1360	0.004891062\\
1520	0.00405469\\
1680	0.002125488\\
1840	0.000672187\\
2000	-0.000392321\\
2160	-0.000868231\\
2320	-0.001877009\\
};

\addplot [color=mycolor3,line width = 1pt, mark=o, mark options={solid, mycolor3}]
  table[row sep=crcr]{%
400	0.012526633\\
560	0.011765698\\
720	0.010529234\\
880	0.009441646\\
1040	0.00801773\\
1200	0.00696847\\
1360	0.00569961\\
1520	0.00482506\\
1680	0.00303887\\
1840	0.00185026\\
2000	0.00066986\\
2160	0.00017651\\
2320	-0.0007617\\
};

\addplot [color=black,line width = 1.2pt, dotted]
  table[row sep=crcr]{%
0  0\\
160 0\\
2400    0\\
};
\end{axis}

\begin{axis}[
set layers,
width=.4\linewidth,
height=.3\linewidth,
at={(-0.1\linewidth,0\linewidth)},
xmin=400,
xmax=2320,
xlabel style={font=\color{white!15!black}},
ymin=0.89,
ymax=1,
ylabel near ticks,
axis y line*=right,
axis x line=none,
y tick label style={mycolor4},
y tick style={mycolor4},
ytick={0.90,0.92,0.94,0.96,0.98,1}, 
yticklabels = {0.90,0.92,0.94,0.96,0.98,1},
title style={at={(0.5,1)}},
title={16-QAM},
legend style={font=\scriptsize, legend cell align=left, align=left, draw=white!15!black}
]
\addplot [color=mycolor4,line width = 1pt, mark=*, mark size = 1pt]
  table[row sep=crcr]{%
400	0.999021644\\
560	0.997167171\\
720	0.993822087\\
880	0.989115708\\
1040	0.98177963\\
1200	0.97681567\\
1360	0.96742639\\
1520	0.95850825\\
1680	0.94246164\\
1840	0.93269759\\
2000	0.92133255\\
2160	0.91576991\\
2320	0.90355249\\
};
\node at (axis cs: 2160,0.9889) [anchor=center] {\large (a)};
\end{axis}
\end{tikzpicture}%
}
    \caption{Normalized net date rate gain of joint CR schemes compared to per-channel estimation for 16-QAM (a), and 64-QAM (b).}
    \label{Fig:results}
\vspace{-1.5\baselineskip}
\end{figure}
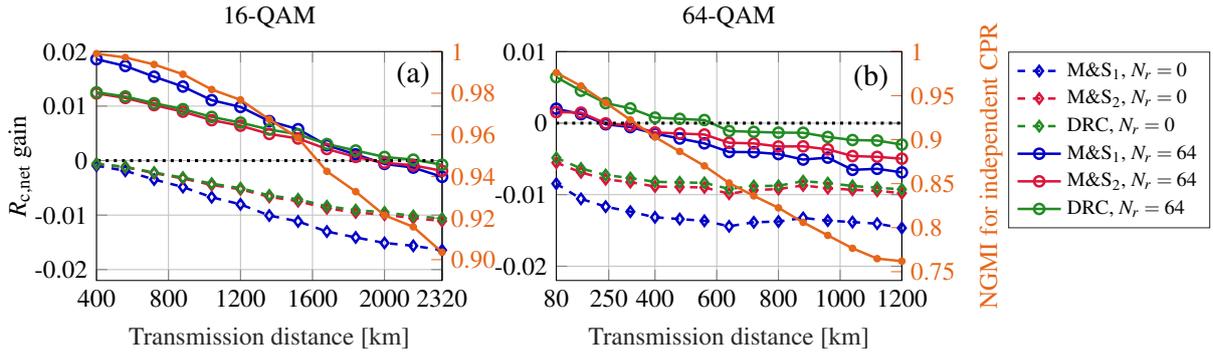

\vspace{-0.5\baselineskip}
We evaluate the performance of the joint CR algorithms in terms of gain in normalized net data rate $R_{\text{c,net}}$ to the baseline independent CR. $R_{\text{c,net}} $ is calculated as $R_{\text{c,net}} = 1/(1+\text{FEC}_{\text{OH}})/(1+\text{POH})$, where $\text{FEC}_{\text{OH}}$ is the forward error correction (FEC) overhead. For a sufficiently small header compared to the frame size, the approximation $\text{POH} \approx 1024/(2^{17}-1024) + \text{POH}_{\text{CR}}$ applies. The $\text{FEC}_{\text{OH}}$ is calculated from the normalized generalized mutual information (NGMI) considering an additional coding gap of 0.07 constant across modulation formats and NGMI values\cite{alvaradoFEC}. The NGMI is averaged over the four channels, assuming joint FEC. The results for 16-QAM and 64-QAM are shown in Fig. \ref{Fig:results} (a) and (b), respectively.

All joint algorithms show a reduction in the net rate for $N_r = 0$ (dashed lines). In this scenario, all CR schemes have equal $\text{POH}_{\text{CR}} = 1/31$, but the lower-complexity PA-CPR does not match the performance of the per-channel PA+DD CPR with optimized window length. The most lightweight scheme M\&S$_1$ is the worst performing, as expected, while M\&S$_2$ and DRC provide similar performance. This behavior suggests that the narrow-spaced PA-CPR partially mitigates CD-induced PN decorrelation, at least for neighboring channels.

The results drastically change when the POH over the secondary channels is optimized. We set $N_r=64$ as a trade-off between a substantial POH reduction and preserving sufficient tracking speed of the additional PA-CPR for the secondary channels. By doing so, joint schemes outperform independent CR for sufficiently short transmission distance. For 16-QAM, M\&S$_1$ is the best-performing algorithm up to $1520 \text{ km}$ transmission due to the largest POH reduction. However, the higher-performance DRC scheme manages to extend the reach for which joint CR outperforms per-channel processing of $240 \text{ km}$, up to $2160 \text{ km}$ at $\text{FEC}_{\text{OH}} \approx 17.5\%$. The benefits of the DRC scheme are larger in relative terms for the more demanding 64-QAM. It outperforms independent CR until $560 \text{ km}$ at $\text{FEC}_{\text{OH}} \approx 23\%$ while M\&S show rate penalties already at half the distance. It is worth noting that M\&S$_2$ is not beneficial for $N_r = 64$, showing the better utilization of the increased POH by the DRC-CPR.

\vspace{-0.5\baselineskip}
\section{Conclusions}
\vspace{-0.5\baselineskip}
We numerically evaluated the performance of joint pilot-aided carrier recovery algorithms in a high-capacity transmission scenario with four 135 GBd signals based on highly-coherent optical frequency combs. We observed that joint schemes and optimization of the pilot overhead can counteract transmission-induced phase noise decorrelation penalties. We reported gains in net information rate for up to 560 km transmission for 64-QAM and 2160 km for 16-QAM over per channel processing, while preserving lower complexity.

\vspace{-0.5\baselineskip}
\section{Acknowledgements}
\vspace{-0.5\baselineskip}
\small
Partially funded by the German Federal Ministry of Education and Research in the project STARFALL (16KIS1418K).

\vspace{-0.5\baselineskip}


\begin{thebibliography}{99} 
\footnotesize 
\bibitem{thin_film_mod} M. Xu, et al., “Dual-polarization thin-film lithium niobate in-phase quadrature modulators for terabit-per-second transmission”, Optica, 2022.

\bibitem{superchannel2015} J. Pfeifle, et al., “Flexible terabit/s nyquist-WDM super-channels using a gain-switched comb source”, Op. Ex., 2015.


\bibitem{ScalingIntegrationChallenges} T. Kobayashi, et al., “Coherent optical transceivers scaling and integration challenges”, Proceedings of IEEE, 2022. 

\bibitem{schroder2019laser} J. Schr{\"o}der, et al., “Laser frequency combs for coherent optical communications”, JLT, 2019.

\bibitem{marin2020performance} P. Marin-Palomo, et al., “Performance of chip-scale optical frequency comb generators in coherent WDM communications”, Op. Ex., 2020.

\bibitem{jorgensen2022petabit} A. J{\o}rgensen, et al., “Petabit-per-second data transmission using a chip-scale micro-comb ring resonator source”, Nature Photonics, 2022.

\bibitem{jasperJOCN} J. M{\"u}ller, et al., “Multi-wavelength transponders for high-capacity optical networks: A physical-layer-aware network planning study”, JOCN, 2023.




\bibitem{alfredsson2019performance} A. F. Alfredsson, et al., “On the performance of joint-core carrier-phase estimation in the presence of intercore skew”, JLT, 2019.

\bibitem{lundberg2020phase} L. Lundberg, et al., “Phase-coherent lightwave communications with frequency combs”, Nature communications, 2020.

\bibitem{lundbergMDPIreview} L. Lundberg, et al., “Frequency comb-based WDM transmission systems enabling joint signal processing”, Applied Sciences, 2018.

\bibitem{ChalmersPowerConsumption} L. Lundberg, et al., “Power consumption savings through joint carrier recovery for spectral and spatial superchannels”, ECOC, 2018. 

\bibitem{neves2021enhanced} M. S. Neves, et al., “Enhanced phase estimation for long-haul multi-carrier systems using a dual-reference subcarrier approach”, JLT, 2021.


\bibitem{Park_sync} B. Park, et al., “A novel timing estimation method for OFDM systems”, IEEE Communications Letters, 2003. 

\bibitem{di2020statistical} G. Di Rosa, et al., “Statistical quantification of nonlinear interference noise components in coherent systems”, Op. Ex., 2020.

\bibitem{alvaradoFEC} A. Alvarado, et al., “Achievable information rates for fiber optics: Applications and computations”, JLT, 2018. 















\end{thebibliography}
\end{document}